\documentstyle[11pt]{article}

\def\seccion#1{\section*{\centerline{\large #1}}}
\def\ital#1{{\it #1}}
\def\ecuacion#1#2#3{$$ #1 \, #2 \eqno (#3) $$}

\begin{document}

\begin{center}
\vspace*{15mm}
{\LARGE {\bf CHESS PURE STRATEGIES ARE PROBABLY CHAOTIC}}\\[3cm]
{\large M. Chaves}\\[3cm]
{\it Escuela de F¡sica\\%
Universidad de Costa Rica\\%
Ciudad Universitaria Rodrigo Facio\\%
San Jos', Costa Rica\\%
e-mail address: mchaves@cariari.ucr.ac.cr}\\[3cm]
\begin{minipage}{125mm}
It is odd that chess grandmasters often disagree in their analysis of
positions, sometimes even of simple ones, and that a grandmaster can hold
his own against an powerful analytic machine such as Deep Blue.
The fact that there must exist pure winning strategies for chess is used to
construct a control strategy function.
It is then shown that chess strategy is equivalent to an autonomous system
of differential equations, and conjectured that the system is chaotic.
If true the conjecture would explain the forenamed peculiarities and would
also imply that there cannot exist a static evaluator for chess.
\end{minipage}
\end{center}
\newpage

\seccion{I.  INTRODUCTION}

Chess grandmasters have great chess playing ability, yet they often disagree
in their analysis of chess states (or positions, as it is said in chess), as
going over the chess literature shows.
Even when performing careful analysis about the same position they often
arrive at differing conclusions.
Rather simple positions with few pieces on the board have been analyzed in
the chess literature for 50 and more years without consensual agreement as to
their theoretical outcome.
How can the analysis be so difficult?

IBM's Deep Blue, a powerful chess playing machine consisting of two
parallel-process tandem supercomputers programmed by a team of experts lead
by team manager C. Tan, [1]-[4] played the world chess champion G. Kasparov
several games in 1996 and 1997 with fairly even results.
Actually, programmer Hsu's estimate back in 1990 of the future machine's
playing strength was 4000 ELO points, far greater than Kasparov's
$\sim$2800 present rating.
In three minutes, which is the game's average pondering time, the machine
could calculate 20 billion moves, enough to for a 24-ply search and an up to
60-ply search in critical tactical lines.
Since grandmasters can calculate just a few moves ahead, it seems strange that
a human could hold his own on the face of such an overwhelming opposition.
Even granting that the grandmaster has access to positional conceptions and
plans, they should be useless against positions calculated so far in advance.

In this paper we offer an explanation for these peculiar situations, based on
the following idea: since chess is a finite 2-person zero-sum game of perfect
information it is strictly determined [5] and thus, for any particular
initial state either White or Black should have a pure winning strategy, or
both sides should have pure drawing strategies.
Now, as we shall discuss, the states of chess can be expressed as points in a
certain 64-dimensional space, so that pure strategies can be pictured as
paths in this space.
But, mathematically speaking, this problem is identical to a
$64^{\rm th}$-order autonomous differential equations problem that is very
likely to be chaotic, so that the paths in the 64 dimensional phase space
must be also chaotic. [6]
Chaos would be an explanation for the peculiar situations just mentioned
because it implies that static evaluators for typical positions in chess
cannot exist.
The horizon problem [7] is exacerbated by the dependence on an evaluator that
cannot be satisfactory.

\seccion{II.  A CONVENTION TO DESCRIBE A CHESS STATE}

To describe a particular state in chess we use a 64 dimensional vector space,
so that to each square of the board we associate a coordinate that takes a
different value for each piece occupying it.
A possible convention could be, for instance, the following: A value of zero
for the coordinate of the dimension corresponding to a square means that
there is no piece there.
For the White pieces the convention would be: a value of 1 for the coordinate
means the piece is a pawn, of 2 means it is a pawn without the right to the
\ital{en passant} move, of 3 that it is a knight, of 4 a bishop, of 5 a rook,
of 6 a queen, of 7 a king, and of 8 a king without the right to castle.
The values for the Black pieces would be the same but negative.

Let us represent the 64-component vector by the symbol $x$.
A vector filled with the appropriate numbers can then be used to represent a
particular state of the game.
We shall call the 64-dimensional space consisting of all the coordinates the
\ital{configuration space} of the game.
The succeeding moves of a pure strategy can be plotted in this space,
resulting in a sequence of points forming a path that we shall call the
\ital{strategy path}.

\seccion{III.  A CONTROL STRATEGY FOR A GIVEN POSITION}

In this section we construct a function that gives, for any arbitrary initial
state of a chess game, the control strategy to be followed by both players.
We begin using the theorem by Zermelo and von Neumann [5] that asserts that
a finite 2-person zero-sum game of perfect information is strictly determined.
For a given initial chess state this implies that either
\begin{itemize}
\item White has a pure strategy that wins,
\item Black has a pure strategy that wins,
\item both are in possession of pure strategies that lead to a forced draw.
\end{itemize}
Consider a certain given initial state of the game where White has a pure
strategy leading to a win.
The two other cases, where Black has the win or both have drawing strategies
can be dealt with similarly and we will not treat them explicitly.

Let the initial chess state be given by the 64-component vector $x_0$, as
defined last section, where we are assuming that White is winning.
The states following the initial one will be denoted by $x_n$, where the
index is the number of plys that have been played from the initial position.
Thus $x_1$ is the position resulting from White's first move, $x_2$ is the
position resulting from Black's first move, $x_3$ is the position resulting
from White's second move, and so on.
Since White has a winning pure strategy, it is obvious that, given a certain
state $x_n$, $n$ even, there must exist a vector function $f$ so that, if 
$x_{n+1}$ is the next state resulting from White's winning strategy,
then $f(x_n)=x_{n+1}$.
On the other hand, if $n$ is odd, so that it is Black's turn, then we define
$f$ to be that strategy for Black that makes the game last the longest before
the checkmate.
Again the strategy would define a function $f(x_n)=x_{n+1}$.
The function $f$ is thus now defined for states with $n$ both even and odd.

A move in chess affects only a few squares, since a piece moves from one
square to another, while the other pieces remain in their original squares.
So it is convenient to introduce the \ital{control strategy vector function}
$g(x_n)=f(x_n)-x_n$, [8] which is much simpler than our previous $f$.
With it we can express the difference between the vectors corresponding to
two consecutive moves as follows:
\ecuacion{g(x_n)=x_{n+1}-x_n}{.}{1}

\seccion{IV.  A CONJECTURE IS ESTABLISHED}

A set of $N$ simultaneous differential equations,
\ecuacion{g(x)={dx \over dt}}{,}{2}
where $t$ is the (independent) time variable, $x \in R^N$ and the $g$ are
known $g: R^N \to R^N$ functions, is called an autonomous system, [6] and
are of frequent occurrence in science.
We will discretize the time variable in (2), as is often done for
computational purposes. [9]
Let the time be within an interval $0 \leq t \leq T$, and then assume it
takes only discrete values $t=0,\Delta t,2\Delta t,...,T$.
After an appropriate scaling of the system one can take the time steps to be
$\Delta t=1$, precisely.
Let the initial condition of the system be $x(0) \equiv x_0$; similarly, let
$x(1) \equiv x_1$, $x(2) \equiv x_2$ and so on.
By taking $N=64$ one can then rewrite (2) in a form that is precisely
identical to (1).

We have shown that the strategy paths of chess in configuration space act as
a 64-dimensional autonomous system.
We have the suspicion that the equivalent autonomous system is chaotic, but
we cannot really prove it, because, of course, we do not really know the
strategy function $g(x)$ for an arbitrary chess initial state and,
considering the complexity of chess, it is hard to imagine we ever could.
(The tree of chess has in the neighborhood of $10^{120}$ moves.)
The following definition is useful to state the coming Conjecture.
We call a state in chess \ital{atypical} if even one of the next is true:
1) it is depleted of pieces; or 
2) there is a forced mate; or
3) one side has overwhelmingly many more pieces than the other does.
And now we establish the

\begin{quote}
\ital{Conjecture. For typical positions $x_0$ in chess the control strategy
vector function results in a chaotic autonomous system.}
\end{quote}
To support this conjecture consider:
\begin{itemize}
\item The usual nonlinear system is chaotic: even simple non-linear ones
usually are.
A chess evaluation function depends on subtle relations between the pieces
so that it could never be linear and, therefore, nonchaotic.
\item The large number of dependent variables (64) makes it more likely that the
autonomous system associated with chess is chaotic.
\end{itemize}
If the position is depleted of pieces, the autonomous system may not be
chaotic.
That is the reason why we have to weaken the conjecture.
If one of the sides has a mate or overwhelmingly many more pieces, the game
will end soon; so these possibilities also have to be excluded.

\seccion{V.  A LEMMA ABOUT STATIC EVALUATORS}

The previous consideration suggests the next lemma that says that practical
chess static evaluators can never be satisfactory.

\begin{quote}
\ital{Lemma. It is not possible for us to build a chess static evaluator
that works with typical chess states.}
\end{quote}
Let us take a typical position, that is, one in which \ital{none} of the
three conditions of an atypical position hold, and see how we could build a
static evaluator for it.
By definition, the static evaluator evaluates on the basis of the position
itself \ital{without recourse to the tree}.
It should be a short program.
We could build an enormous program in a practically infinite computer that
plays all possible chess games starting from a position and decides how good
the position is, \ital{but this is not what a static evaluator is}.
The evaluation of the state has to be done using heuristics, that is, it has
to use simple rules that say how good a state is on the basis of the
positions of the pieces.
But this is not possible if chess is chaotic because we know that the
smallest difference in the positioning of a piece will lead to completely
diverging paths in configuration space, that is, to wholly different states.
Therefore there would have to be as many rules as typical states there are,
and the resulting program would far too long to be a static evaluator. 

We referred in the Introduction to the horizon effect.
This is a problem that occurs in game programming when the computer quits
the search in the middle of a critical tactical situation and thus it is
likely that it gets an incorrect evaluation from the heuristics.
Since the Lemma assures us that practical chess static evaluators cannot be
very good, we arrive at the conclusion that the horizon problem is
exacerbated in chess.
The reason is that, independently of how deep the search is, at the tips of
the strategy tree there is going to be an evaluator that does not work
satisfactorily.
On the other hand, going deeper will always be helpful since by explicitly
searching over more plys we lighten the evaluator's job, so to speak.
We shall call this effect of always being able to go deeper in the search but
at the same time having the evaluator not evaluating correctly at least some
of the states the \ital{receding horizon effect}.

Chess has the characteristic that states tend to get progressively simpler
due to piece depletion.
Eventually all initially typical states will cease to be so as we follow
the tree.
Thus chess is chaotic only temporarily.
To be able to state nicer theorems the rules of chess would have to be
changed so that the pieces keep returning to the board.

\seccion{VI.  CONCLUSION AND COMMENTS}

We have seen that it is likely that the pure strategy paths of chess in
configuration space follow chaotic paths, and that this implies that
practical static evaluators cannot be very good.
As a result the horizon problem is enhanced; this is the situation we call a
receding horizon.
We have commented that grandmasters often disagree in their analysis of a
particular position.
Sometimes discussions about a position go on for many years.
On the basis of our results, this situation can be understood as follows:
a grandmaster can only search a few plys, and then has to evaluate the
resulting position statically.
But since we have seen that static evaluators cannot be too good, we obtain
the general result that human analysis cannot be very good, either.

The reason why a machine such as Deep Blue is not far stronger than a human
has to do again with the problem of building a static evaluator.
Even though the machine searches many more plys than the human does, at the
end it has to use a deficient static evaluator.
Now, the fact that Deep Blue has grandmaster strength is telling us that the
human mind has a far better static evaluator than Deep Blue.
But this only makes a difference because chess is chaotic, otherwise, since
the machine can see ahead many more plys than the human, the fact the human
has a better evaluator would not count.
Human intuition has a chance playing chess against supercomputers because
the game is chaotic.
It should be possible to program a supercomputer to play a nonchaotic game
perfectly.

How to improve evaluators? The human mind is a neural network; perhaps neural
networks could be tried to as static evaluators.
They can be made to learn by playing many games with different settings.
Perhaps the machine can be programmed so that it learns, as was done for
another game, link-five. [10]
Another different approach to the problem of improving evaluators would be
to use the positional concepts that have been developed by grandmasters over
the years.
We are not referring simply to the heuristics of a static evaluator based on
centralizing pieces, overexposed kings and the like, as the ones used in the
tips of the tree in Type A control strategy, or to decide which branches are
going to be inspected in Type B, [11] but instead to what chessmasters call
``superior strategy", examples of which are ``playing with a plan" and
``accumulation of small advantages". [12]
At any rate it would seem, on the basis of our considerations, that the
introduction of advanced heuristics is unavoidable to assure a continuous
rapid strengthening of chess playing programs.
The problems involved belong in the specific area of artificial intelligence.

\seccion{ACKNOWLEDGEMENT}

I wish to thank Dr. Walter Fern ndez, head of the Laboratorio de
Investigaciones Atmosf'ricas y Planetarias, for kindly allowing me to become
a rather frequent user of the computers there during this past year.

\end{document}